\documentclass[10pt, conference, compsocconf]{IEEEtran}
\usepackage{amsmath}
\usepackage{algorithm}
\usepackage{algorithmic}
\usepackage{color}
\usepackage{stfloats}
\usepackage{supertabular,booktabs}
\usepackage{graphicx}
\usepackage{bm}
\usepackage[colorlinks,linkcolor=red,anchorcolor=blue,citecolor=green,CJKbookmarks=True]{hyperref}
\begin{document}
\title{Constructing Financial Sentimental Factors in Chinese Market Using Natural Language Processing}
\author
{\IEEEauthorblockN{Junfeng Jiang \IEEEauthorrefmark{1}\IEEEauthorrefmark{2},Jiahao Li\IEEEauthorrefmark{1}\IEEEauthorrefmark{2}}
\IEEEauthorblockA
{
\IEEEauthorrefmark{1}Likelihood Technology\\
}
\IEEEauthorblockA
{
\IEEEauthorrefmark{2}Sun Yat-sen University\\
}
$ $\\
$\{Jiangjf6,lijh76\}@mail2.sysu.edu.cn$
}

\maketitle
\begin{abstract}
In this paper, we design an integrated algorithm to evaluate the sentiment of Chinese market. Firstly, with the help of the web browser automation, we crawl a lot of news and comments from several influential financial websites automatically. Secondly, we use techniques of Natural Language Processing(NLP) under Chinese context, including tokenization, Word2vec word embedding and semantic database WordNet, to compute Senti-scores of these news and comments, and then construct the sentimental factor. Here, we build a finance-specific sentimental lexicon so that the sentimental factor can reflect the sentiment of financial market but not the general sentiments as happiness, sadness, etc. Thirdly, we also implement an adjustment of the standard sentimental factor. Our experimental performance shows that there is a significant correlation between our standard sentimental factor and the Chinese market, and the adjusted factor is even more informative, having a stronger correlation with the Chinese market. Therefore, our sentimental factors can be important references when making investment decisions. Especially during the Chinese market crash in 2015, the Pearson correlation coefficient of adjusted sentimental factor with SSE is 0.5844, which suggests that our model can provide a solid guidance, especially in the special period when the market is influenced greatly by public sentiment.
\end{abstract}


\IEEEpeerreviewmaketitle
\section{Introduction}

Natural language processing, as one of the most promising fields in machine learning, has achieved great development recently and has been used in lots of aspects in the society. Many researches also implemented the technique of NLP in financial market. The difficulty when applying NLP is that the natural language is not a kind of structural data. Finding a way to process such kind of non-structural data is the main focus of NLP. A lot of models have been demonstrated to do well in turning natural language data into numerical data, which is more tractable. With the implementation of these models, it becomes possible and easier to make use of natural language data.

Some models are based on the idea of Naive Bayesian[\hyperref[ref 1]{1}]. The logic behind these models is: words that show the same kind of sentiment will appear simultaneously more frequently. These models usually select some words as the label words. By analyzing the words appeared in a large amount of texts and researching on the relationship between the frequency of these label words and the frequency of other words, it becomes possible to cluster words. For any given texts, it is able to use these words to evaluate the sentiment behind. Researches have proved this kind of methods can successfully evaluate the sentiment of texts like twitter or news. By taking advantage of the sentiment, investors can make appropriate investment decisions.

However, this kind of methods have their own limitations. The main is that they just focus on a few words. Some new words that show the similar sentiment but do not appear frequently will be ignored. Sometimes unfortunately, these words do play a significant role when analyzing the sentiment of a text. The lost of information can have great damage to the accuracy of the evaluation.

This study is aiming to put eyes as many words as possible to analyze sentiment. The specific steps of this study are as followed:

\begin{itemize}
\item Download news from several influential financial websites automatically.
\item Make a pre-treatment on the news we crawl from the Internet.
\item Find a way or some algorithms to analyse this per-treated text data, and finally compute the sentimental factor of each day with the news at that day.
\item Choose a proper criterion to analyze the correlation between the sentimental factor and the market trend, and judge whether our factor is useful in financial investment.
\end{itemize}

The code we use are open source in Github.\footnote{\href{https://github.com/Coldog2333/Financial-NLP}{https://github.com/Coldog2333/Financial-NLP}}

The remainder of this paper is organized as follows: Section 2 describes the research background and related works of Jieba, Word2vec and WordNet. Section 3 shows the methodology and the data we use in analysis. Section 4 contains the experimental result and discussion. Finally, in Section 5, we proposed our conclusions.

\section{Related Works}
\subsection{\underline{Jieba}}
The tokenization of Chinese is much more complicated than English. To tokenize English words, we just need to split words in sentence by blank or punctuation. Chinese doesn't have blank between words. An additional step of tokenization is, therefore, needed.

Jieba(Chinese for "to shutter") Chinese text tokenization is a Chinese word tokenization module. The algorithm of Jieba is probability language modeling. It generates a trie tree based on a dictionary transcendentally and also calculate the frequency of words in the dictionary. When dealing with the sentence that is needed to be tokenized, it generates a DAG(Directed Acyclic Graph) to record every possible tokenization.A DAG is a dictionary, where the keys are the starting position of a word in the sentence and the values are lists of possible ending position.

For every possible words in the DAG, Jieba calculates their probability based on the transcendental dictionary. Then it find the path with the largest probability from the right side of the sentence to the left side. This largest probability path gives us the most possible tokenization.

In the case where the sentence includes words that are not in the dictionary, Jieba uses HMM(Hidden Markov Model) and Viterbi algorithm to tokenize. Every character has four conditions based on its possible condition in a word: B(Begin), M(Middle), E(End) and S(Single). The process of tokenizing words not in the dictionary is based on their conditions mainly. With three probability tables from the training of a large amount of texts, Jiaba then applies Viterbi algorithm to calculate the most possible condition of a word and uses the conditions chain to tokenize.

\subsection{\underline{Word2vec}}

In 2013, Google published a powerful tool named word2vec[\hyperref[ref 2]{2}]. It contains two models, one is the Skip-gram, another is Continuous bag of words(CBOW). With the word2vec model, we can turn a specific word into a calculable numeric vector. To speak of, moreover, it can well express the degree of similarity and analogy between two different words.

Since word2vec have been published, it is widely applied in Natural Language Processing, and its original models and training methods also enlighten many word embedding models and algorithms on the following days. Now, we introduce the word2vec model with an English example.

\subsubsection{\underline{Skip-gram}}
In Skip-gram, we focus on one word, and use it to predict which words will appear around it.

For example,"the boy adores that girl", we can achieve five background words like "the", "boy", "adores", "that", "girl" easily because we have blanks between every two words. Let "adores" be the center word, and set window size equals to 2, then, in Skip-gram, what we are interested in is the conditional probabilities of each background word under the given center word, where the background words is apart from the center word in two words. That is the mainly idea of Skip-gram. Let's describe the Skip-gram model in a strict mathematical language.

Assume that size of the set of dictionary index \textit{D} is $\big|$\textit{D}$\big|$, and denoted as \textit{D}=\{1,2,...,$\big|$\textit{D}$\big|$\}. Given a text sequence with the length of \textit{T}, and the $t^{th}$ word denoted as $w^{(t)}$.When window size equals to m, Skip-gram requires that we should maximize the total of all conditional probabilities of each background word that is apart from the center word in \textit{m} words under arbitrary center word.

\begin{equation}
\prod_{t=1}^{T}\prod_{-m \leq j \leq m, j\neq 0, 1\leq t+j \leq \big|T\big|}P(w^{(t+j)}\big|w^{(t)})
\end{equation}

So, the likelihood function is,

\begin{equation}
\sum_{t=1}^{T}\sum_{-m \leq j \leq m, j\neq 0, 1\leq t+j \leq \big|T\big|}log P(w^{(t+j)}\big|w^{(t)})
\end{equation}

Maximizing the likelihood function above minimize the following loss function,

\begin{equation}
-\frac{1}{T} \sum_{t=1}^{T}\sum_{-m \leq j \leq m, j\neq 0, 1\leq t+j \leq \big|T\big|}log P(w^{(t+j)}\big|w^{(t)})
\end{equation}

Denote the vectors of center words and background words with \textbf{v} and \textbf{u}, that is, as for a word with index \textit{i}, $\textbf{v}_{i}$ and $\textbf{u}_{i}$ are the vectors when it is as center word and background word. And the parameters of model we want to train are the two kinds of vectors of every words.

In order to implement the model parameters into loss function, we should express the conditional probabilities of background word under given center word with model parameters. Assume that generating each background words is independent mutually when center word is given, then as for the center word $w_{c}$ and the background word $w_{b}$, b, c are the indexes of them in the dictionary. Such that, the probability of generating background word $w_{b}$ under the given center word $w_{c}$ can be defined by softmax function, as

\begin{equation}
P(w_{b}\big|w_{c})=\frac{exp(\textbf{u}_{b} ^{T}\textbf{v}_{c})}{\sum_{i\in D}exp(\textbf{u}_{i}^{T}\textbf{v}_{c})}
\end{equation}

With derivation, we achieve the gradient of the conditional probability above,

\begin{equation}
\frac{\partial logP(w_{b}\big|w_{c})}{\partial \textbf{v}_{c}}=\textbf{u}_{b}-\sum_{j\in D}\frac{exp(\textbf{u}_{j} ^{T}\textbf{v}_{c})}{\sum_{i\in D}exp(\textbf{u}_{i}^{T}\textbf{v}_{c})}\textbf{u}_{j}
\end{equation}

Namely,

\begin{equation}
\frac{\partial logP(w_{b}\big|w_{c})}{\partial \textbf{v}_{c}}=\textbf{u}_{b}-\sum_{j\in D}P(w_{j}\big|w_{c})\textbf{u}_{j}
\end{equation}

Then, we can solve this by Gradient Descent or Stochastic Gradient Descent iteratively, and finally achieve the word vectors $v_{i}$ and $u_{i}$, \(i=1,2,...,\big|D\big|\) of every single words when it is as center word and background word, when the loss function reaches to minimum.

If the length of text sequence \textit{T} is too long, we can sample a rather short subsequence randomly to calculate the loss about this subsequence in each epoch, in order to find out an approximate solution.

In general, we will use the central word vector of Skip-gram as the word vector of each word in natural language processing application.

\subsubsection{\underline{Continuous Bag of Words}}

CBOW is similar to Skip-gram, this model predicts the central word with the background words around it in a text sequence. For example, "the boy adores that girl", we can achieve five background words like "the", "boy", "adores", "that", "girl". Let "adores" be the central word again, and set window size equals to 2, then, in CBOW, what we are interested in is the conditional probabilities of generating the given central word under all the background words which are apart from the central word in two words. That is the mainly idea of CBOW.

Assume that size of the set of dictionary index \textit{D} is $\big|\textit{D}\big|$, and denoted as \textit{D}=\{1,2,...,\big|\textit{D}\big|\}. Given a text sequence with the length of \textit{T}, and the $t^{th}$ word denoted as $w^{(t)}$.When window size equals to m, CBOW requires that we should maximize the total of all conditional probabilities of generating the arbitrarily given central word under all the background words which are apart from the central word in m words.

\begin{equation}
\prod_{t=1}^{T}P(w^{(t)}\big|w^{(t-m)},...,w^{(t-1)},w^{(t+1)},...,w^{(t+m)})
\end{equation}

where \textit{m} is the window size, and we should insure that (t-m+j)$\in$[1,\big|\textit{T}\big|],j$\in$[0,2m].

Therefore, the likelihood function is,

\begin{equation}
\sum_{t=1}^{T}logP(w^{(t)}\big|w^{(t-m)},...,w^{(t-1)},w^{(t+1)},...,w^{(t+m)})
\end{equation}

Maximizing the likelihood function above minimize the following loss function,

\begin{equation}
-\sum_{t=1}^{T}logP(w^{(t)}\big|w^{(t-m)},...,w^{(t-1)},w^{(t+1)},...,w^{(t+m)})
\end{equation}

We still use the notation when we discuss Skip-gram model. Now, as for the central word $v_{c}$ and its background words $w_{b0}$,$w_{b1}$,...,$w_{b\cdot 2m}$, such that, the probability of generating the given central word $w_{b}$ under all the background words $w_{b1}$,$w_{b2}$,...,$w_{b\cdot 2m}$ can be defined by softmax function as,

\begin{equation}
\begin{aligned}
& P(w_{c}\big|w_{b0},w_{b1},...,w_{b\cdot 2m}) \\
& = \frac {exp( \frac{\textbf{v}_{c}^{T}(\textbf{u}_{b0}+\textbf{u}_{b1}+...+\textbf{u}_{b\cdot 2m})}{2m})} {\sum_{i\in D}exp(\frac{\textbf{v}_{i}^{T}(\textbf{u}_{b0}+\textbf{u}_{b1}+...+\textbf{u}_{b\cdot 2m})}{2m}) } \\
\end{aligned}
\end{equation}

With derivation, we achieve the gradient of the conditional probability above,

\begin{equation}
\begin{aligned}
& \frac{\partial logP(w_{c}\big|w_{b0},w_{b1},...,w_{b\cdot 2m})}{\partial \textbf{u}_{bi}} \\
& =\frac{1}{2m} (\textbf{v}_{c}-\sum_{j\in D} \frac{exp( \frac{\textbf{v}_{c}^{T}(\textbf{u}_{b0}+\textbf{u}_{b1}+...+\textbf{u}_{b\cdot 2m})}{2m})} {\sum_{i\in D}exp(\frac{\textbf{v}_{i}^{T}(\textbf{u}_{b0}+\textbf{u}_{b1}+...+\textbf{u}_{b\cdot 2m})}{2m}) }\cdot \textbf{v}_{j}) \\
\end{aligned}
\end{equation}

Namely,

\begin{equation}
\begin{aligned}
& \frac{\partial logP(w_{c}\big|w_{b0},w_{b1},...,w_{b\cdot 2m})}{\partial \textbf{u}_{bi}} \\
& = \frac{1}{2m} (\textbf{v}_{c}-\sum_{j\in D} P(w_{c}\big|w_{b0},w_{b1},...,w_{b\cdot 2m})\cdot \textbf{v}_{j}) \\
\end{aligned}
\end{equation}

As the same of Skip-gram, we can also solve this by Gradient Descent or Stochastic Gradient Descent iteratively, and finally achieve the word vectors $\textbf{v}_{i}$ and $\textbf{u}_{i}$ (\textit{i}=1,2,...,\big|\textit{D}\big|) of every single words when it is as center word and background word, when the loss function reaches to minimum.

If the length of text sequence T is too long, we can sample a rather short subsequence randomly to calculate the loss about this subsequence in each epoch, in order to find out an approximate solution.

In general, we will use the background word vector of CBOW as the word vector of each word in natural language processing application.

\subsection{\underline{WordNet}}

WordNet is a large lexical database of English. In WordNet, synsets are interlinked by means of conceptual-semantic and lexical relations. The main relation between words in WordNet is synonym[\hyperref[ref 3]{3}]. By using a network to show the relation between words, WordNet helps us find synonym of words and also shows how two much words are similar with each other in the perspective of meanings.

\begin{figure}[ht]
\centering
\includegraphics[scale=0.2]{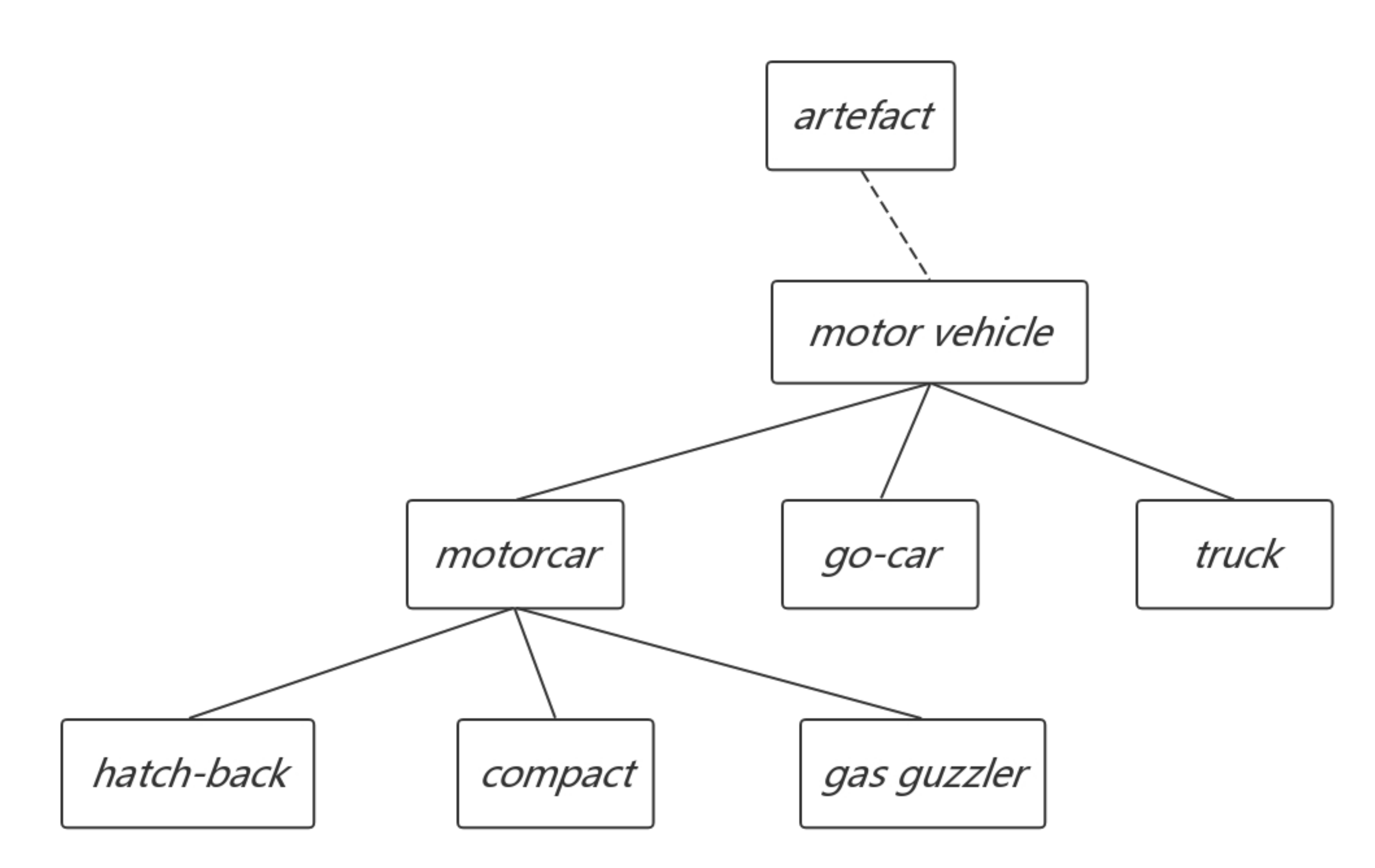}
\caption{An Example of the Structure of WordNet[\hyperref[ref 4]{4}]}
\end{figure}

Having the properties above makes WordNet more reliable to analyze the sentiment. However, when using NLP under Chinese context, we should also translate Chinese to English at first. Thanks to the contributors of Chinese Open Wordnet (COW), as for a Chinese word, we can conveniently find the corresponding meaning in English, namely, we can find the corresponding node of this Chinese word in WordNet. COW is a large scale, freely available, semantic dictionary of Mandarin Chinese inspired by WordNet[\hyperref[ref 5]{5}]. It has the same structure and principles as WordNet but based on Chinese. It contains 42,315 synsets, 79,812 senses and 61,536 unique words and the construction is still ongoing. Our research mainly uses COW to compute the Senti-score of each single words.

\section{Methodology}
\subsection{\underline{Data Mining}}

In order to gain plenty of financial news, we use selenium\footnote{\href{https://www.seleniumhq.org/}{https://www.seleniumhq.org/}} to crawl the news from network. We collect the historical data automatically and use them to compute historical Senti-scores.

At first we crawl from Xueqiu, bjzq (Beijing Securities website) and Chinastock. But later we find that Chinastock contains articles from a much longer time and broader areas. The sources of Chinastock already include many financial websites. As a result, we choose Chinastock to do text data crawling.

In order to do back testing more conveniently, we sort the text data by date.

\subsection{\underline{Pre-treatment}}

Chinese is one of typical isolated languages. Instead of using inflections, Chinese uses isolated function words and various word order to express grammar. Compared with English, Chinese has its uniqueness. Unlike English, Chinese does not use blank to tokenize words. Therefore, before doing natural language processing over Chinese, we need to have one more tokenization step, which turns an article into a bag of words. It is a significant difference between Chinese and English natural language processing. For any given piece of news or article, we use Jieba for pre-treatment. Besides turning the whole article into a bag of words, we also need to eliminate the stop words which are not important and will affect our analysis.

However, when we train our model with the corpus, we do not remove the stop words from the corpus because the lost of stop words in such a case will affect the generation of the proper word vector. We will make a concrete analysis of this question in \textbf{Text Analysis} below.

Besides, we also need to define some symbols like numbers and punctuation that needed to be removed in order to lessen the amount of calculation and noise.

\subsection{\underline{Text Analysis}}
\subsubsection{\underline{Morphological Similarity}}

At the beginning, we should train our word2vec model achieving more word vectors as much as possible (Theoretically we can achieve word vectors of all words as long as we train word2vec with a large enough corpus, but in fact, we cannot always have such a corpus). For the generality, we choose zh-wiki as our training corpus.

It should be noted that we cannot remove stop words on training corpus because stop words can be significant describing a specified central word.

Set an appropriate large length of word vector in order to make words linearly separable in the hyper plane as far as possible. And we can set a proper threshold of frequency to omit some unfamiliar words, saving memory without losing any vital information.

After training, we achieve a word2vec model, meanwhile we get to know the word vectors of all words. We should just look up a word vector of a word in the model(like a dictionary).

Now, we can calculate the morphological similarity of two words with their cosine distance. For example, let's consider two words $w_{1}$ and $w_{2}$, and normalize as $w_{1}^{'}$ and $w_{2}^{'}$, so

\begin{equation}
distance = \frac{w_{1}\cdot w_{2}}{\|w_{1}\|\|w_{2}\|} = w_{1}^{'}\cdot w_{2}^{'}
\end{equation}

which indicates the morphological similarity of two words $w_{1}$ and $w_{2}$.

\textbf{Note}: Though now we can estimate the similarity of two words with word2vec, what we achieve above is only the similarity morphologically. That is to say, we can just find out which words are similar with a specified word morphologically, but do not know their meanings. Let's take an easy example, consider two words 'increase' and 'decrease', we can often read such a sentence 'The .DJI increases by 5 percent today' on the financial news, and you will find it certainly possible to exchange 'increase' into 'decrease' in this sentence without any difficulty. As known in \textbf{2.2}, the word vectors of this two words will approximately be the same, which may make us confused in determine their respective Senti-scores. But, what deserves to be mentioned is that word2vec does help us to find out some words familiar with a given word in a manner.

Facing this embarrassing situation, we propose a relatively well method in \textbf{\underline{\textit{D.Senti-score of Words}}}.

\subsubsection{\underline{Semantics Similarity}}

As mentioned in \textbf{2.3}, WordNet uses trees to record words. The structure of trees defines distances naturally. When computing the semantic similarity of words using WordNet, we use the shortest path linking the two nodes representing the words to compute. Take the reciprocal of the shortest path of the two nodes as the similarity. The similarity between a word and itself is 1. We also define the similarity between two words is 0 if there is no path linking them. With definitions above, the semantics similarity we compute will be a value between 0 and 1. The larger the value is, the more similar the two words are semantically.

\subsection{\underline{Senti-score Computation}}
\subsubsection{\underline{Sentiment Lexicon}}

Generally in sentiment analysis, we will build a sentiment lexicon to tell our model that which words are positive or negative. It inspires us to define our specific sentiment lexicon so that our model can output the sentiment which reflects financial market.

Then, we define a sentiment lexicon as a small tally set to evaluate the sentiment of a single word. The idea is that it is possible to reflect the sentiment of a word by calculating the similarity of this word with the word in sentiment lexicon, which we initialize its sentiment. We choose 100 words that appear in financial news frequently and also have specific sentiment as our label words. On the other hand, in order to make the computation more fair, we choose 50 words that have positive attitude toward the market (positive words) and the other 50 have the opposite sentiment (negative words). Before computing the Senti-score of a piece of new, we firstly compute both the morphological similarity and the semantics similarity with the label words of every single word.

\hyperref[Table 1]{Table 1} provides some words in our sentiment lexicon as an example. The whole sentiment lexicon displays in our codes on our GitHub.\footnote{\href{https://github.com/Coldog2333/Financial-NLP}{https://github.com/Coldog2333/Financial-NLP}}
\begin{table}[!ht]
    \centering
    \begin{tabular}{c|c}
        \hline
        \textbf{Positive} & \textbf{Negative}\\
        \hline
        {bullish} & {bearish} \\
        {climb} & {fall} \\
        {surge} & {slump} \\
        {...} & {...} \\
        {hortation} & {sanction} \\
        \hline
    \end{tabular}
    \caption{Sentiment Lexicon}\label{Table 1}
\end{table}

\subsubsection{\underline{Senti-score of Words}}

With the computation above, we will have a vector of 200 dimensions for every word. The first 100 dimensions are the Word2vec similarities with the 100 words in the sentiment lexicon. They show the morphological similarity with the label words in sentiment lexicon. The last 100 dimensions, on the other hand, are the WordNet similarities with the 100 words in the sentiment lexicon, which represent the semantic similarity with the label words. Since the words in the sentiment lexicon show attitudes toward the market, based on the similarities above, we can evaluate the attitude of a specific word, namely the sentiment of it. To represent the sentiment quantitatively, we define a value called Senti-score.

In our research, we use the similarity vector to compute the Senti-score of every word. In specific, the process includes the following steps: (1)Use the Word2vec similarity and the WordNet similarity respectively. Then use collaborative filtering to classify it as positive word or negative word. (2)Use Word2vec similarity to compute the Senti-score. The specific process and the reasons behind are as followed.

For a word, when we consider the similarities of it with those words in the sentiment lexicon, there are two kinds of similarities we should think about. The first one is the morphological similarity. The second one is the semantic similarity. Collaborative filtering can help us to find the several most similar words morphologically and semantically. First of all, we use the first 100 dimensions, which are the Word2vec similarities, to find the top \textit{n} similar words. We find them by picking up these \textit{n} words that have the largest value in the first 100 dimensions. The words we find in this way will be the \textit{n} words that are the most morphologically similar to the target word we need to compute. Then we compare the WordNet similarities of these \textit{n} words, which are in the last 100 dimensions. From these \textit{n} words, we pick up top\textit{m} of them that have the largest WordNet similarities. These \textit{m} words will be the top \textit{m} similar to the target word both morphologically and semantically.

With these \textit{m} words we can judge whether the target word is positive or negative. However, it is not enough if we just label it as +1 or -1. The reason is that the degree of positive or negative can be different for the same kind of words. A positive word can be more positive than another positive word. Therefore, we need to define a score to measure how positive or negative the word is. The method is, we firstly define the scores of positive words and negative words in sentiment lexicon are +1 and -1. Then we use Word2vec similarities of the \textit{m} words with a target word as the weights, and calculate the weighted average to be the Senti-score of a target word.

A problem we have to face is that the words included in the COW is far less than words in the Word2vec model we trained. In some cases, the last 100 dimensions of the word cannot be computed. In these cases, we use the Word2vec similarity only to evaluate. For these words, in the second step of collaborative filtering, we cannot find the top \textit{m} Semantically similar words from the top \textit{n} morphologically similar words. Therefore, we can just find the top\textit{m} morphologically similar words from top \textit{n} morphologically similar words. For the same reason, we still compute the Senti-score of every word weighted by Word2vec similarity.

Up to now, we can compute the Senti-score of every single word.

\subsubsection{\underline{Senti-score of Articles and Sentimental Factor}}

Use the Senti-scores of words above, we can compute the Senti-score of every piece of news. By adding up all the Senti-score of news in a day, we will get the sentimental factor we need. Details are as follows.

After the pretreatment, we will get a word bag. It is straight-forward to compute the similarity vector of each word in the word bag and use these vectors to compute the Senti-scores. However, it is not efficient and not necessary. Too many words will be computed for many times. The fact is, if we compute the Senti-socres of the most commonly used words in advance, we can get the Senti-score of an article much quicker and much more efficient.

We use over 50,000 common words in \emph{The Common Vocabulary of Modern Chinese} published by Commercial Press and pick up words that are not in these over 50,000 words in 3,000 pieces of news. Collect these words and we get a common word set of around 100,000 words. We compute the Senti-scores of words in common word set in advance.So that when we compute the Senti-score of an article, we will just need to look up the common word set. Some words may be ignored if we use this method, but our experiment results show that words not in the common word set will only be around 5\% of words in an article. The influence of ignoring the 5\% is just a drop in the bucket.

Use the method above we can compute Senti-score of news. We compute the average Senti-score of a day and use it as the sentimental factor that day. The sentimental factor computed in this way can show the sentiment of the market effectively.

\subsubsection{\underline{Adjustment of model}}

Until now, we have computed the Senti-score of every single day. When we make a time series analysis with the Senti-score and the market index on the same day, we find that though market index change smoothly, the Senti-score come to a violent fluctuations, which is inconsistent with reality. Therefore, we have to adjust the model before applying it. Then we smooth the Senti-score, that is, when evaluating the Sentiment scores of one day does not mean we should use the Senti-score of that day directly, but taking the average of the Senti-score of a period of time. It's reasonable because of the timeliness of information. As we know, the public sentiment of a day will not only influence the market at the same day, but also influence the market in the following days. Next, we will compare the adjusted model with the original model in the following experiment.

\subsection{\underline{Correlation Analysis}}
We need a criterion to evaluate the efficiency of the sentimental factor. The main criterion is to do a linear regression using the sentimental factor and the market trend[\hyperref[ref 7]{7}]. Assume that the market sentiment has positive correlation with the market trend, then the significance of the regression result will be able to show the efficiency of the sentimental factor. What's more, we can also evaluate their correlation of them with Pearson correlation coefficient below.

\begin{equation}
\rho_{X,Y} = \frac{E[X\dot Y]-E[X]E[Y]}{\sigma_{X}\dot \sigma_{Y}}
\end{equation}

\section{\underline{Baseline}}

\textbf{Random.} We randomly create some series of real numbers as a random factor to compare with our sentimental factor. The one is generated by uniform distribution, which scales from the minimum to the maximum of our sentimental factor. Another is generated by normal distribution, which has the mean and the standard deviation same as our sentimental factor.

\textbf{Temperature.} Edward M. Saunders Jr.[\hyperref[ref 8]{8}] proposed that weather can also influent the financial market. To see whether our complex sentimental factor can outperform than the simply found factor, we compare it with our sentimental factor as a baseline. We download climatic data from NNDC.\footnote{\href{https://www7.ncdc.noaa.gov/CDO/cdoselect.cmd}{https://www7.ncdc.noaa.gov/CDO/cdoselect.cmd}} We especially choose the climatic data in Shanghai and Guangzhou.

\section{\underline{Experimental Results and Discussion}}

We calculate the standard sentimental factor of 1379 market days from 2012/11/6 to now, and make a correlation analysis about the standard sentimental factor and some market indexes like SSE and SZSE.

Linear regression is carried out with them, and we can see that all of the coefficients successfully pass significance test in \hyperref[Table 2]{Table 2} and \hyperref[Table 3]{Table 3}.

\begin{table}[h]
\centering
\caption{The regression result of the standard sentimental factor with SSE}\label{Table 2}
\begin{tabular}{c|c}
\hline
{standard sentimental factor} & 11138.27 ***\\
{p} & (3.48e-12) \\
\hline
\end{tabular}
\end{table}

\begin{table}[h]
\setlength{\abovecaptionskip}{1.mm}
\setlength{\belowcaptionskip}{-5.mm}
\caption{The regression result of the standard sentimental factor with SZSE}\label{Table 3}
\centering
\begin{tabular}{c|c}
\hline
{standard sentimental factor} & 39796.6 ***\\
{p} & ($<$2e-16) \\
\hline
\end{tabular}
\end{table}

This time the Pearson correlation coefficient is 0.18731 (with SSE), which states the standard sentimental factor has a weak dependence with the market index. However, when we adopt the adjusted sentiment factor and apply a similar analysis, we can find that the Pearson correlation coefficient is 0.26119, which is improved a lots.

Especially, we do another experiment during the period of the China stock market crash, from 2015/02/11 to 2015/09/11. We compute the sentimental factor of 139 market days of this year, and similarly make a correlation analysis about the sentimental factor with SSE and SZSE.

This time the Pearson correlation coefficient is 0.36284, which states the Sentiment factor has a medium dependence with the market index. Moreover, when we adopt the adjusted Sentiment factor and apply a similar analysis, we can find that the Pearson correlation coefficient is 0.58815, which shows a higher dependence. What's more, when we make a time series analysis, we can see that especially during the period of the stock market crash, the sentimental factor moves almost the same as the market index.

\begin{figure}[ht]
\centering
\includegraphics[scale=0.12]{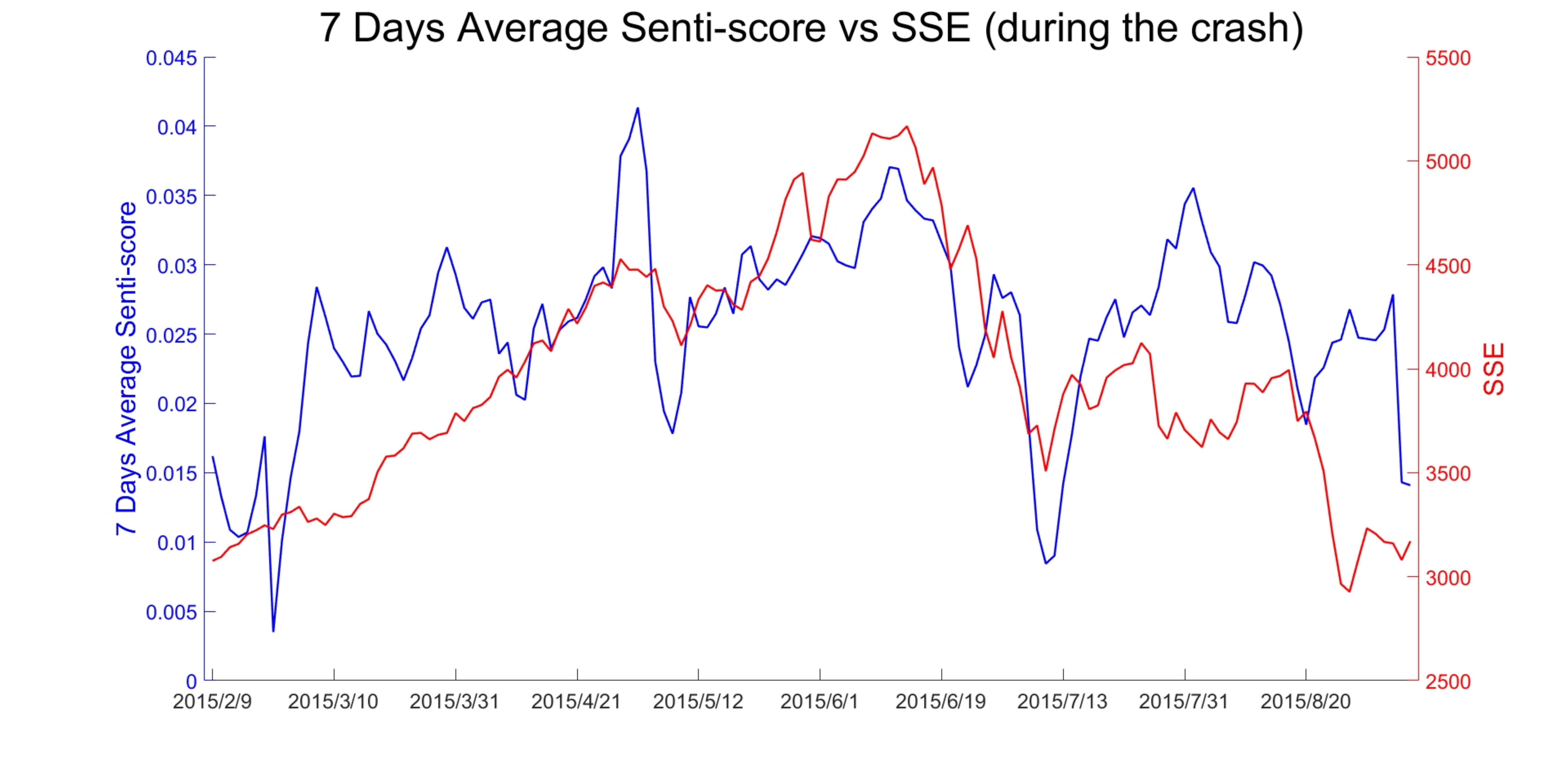}
\caption{Time Series of 7 Days Average Senti-score and SSE during the crash}
\end{figure}

Meanwhile, we apply the similar correlation analysis about the random factor and temperature with SSE and SZSE, and the complete result displayed on \hyperref[Table 4]{Table 4} and \hyperref[Table 5]{Table 5}. For reducing random errors, when we applying correlation analysis on random factor, we create 1000 series of real numbers and compute the average of these 1000 Pearson correlation coefficient as the final result.

\begin{table}[!ht]
    \setlength{\abovecaptionskip}{-5.mm}
    \setlength{\belowcaptionskip}{-0.cm}
    \centering
    \begin{tabular}{ccc}
        \hline
        \textbf{Pearson correlation coefficient}\\
        \hline
        {} & {SSE} & {SZSE} \\
        \hline
        {random from uniform} & {-0.00049399} & {-0.0005597} \\
        {random from normal} & {0.00017451} & {0.00016782} \\
        {temperature} & {-0.025135} & {-0.063723}\\
        {standard sentimental factor} & {0.18731} & {0.22595}\\
        {adjusted sentimental factor} & {\textbf{0.26119}} & {\textbf{0.28472}}\\
        \hline
    \end{tabular}
    \caption{Pearson correlation coefficient from 2012/11/6 to 2018/8/17}\label{Table 4}
\end{table}

\begin{table}[!ht]
    \setlength{\abovecaptionskip}{-5.mm}
    \setlength{\belowcaptionskip}{-0.cm}
    \centering
    \begin{tabular}{ccc}
        \hline
        \textbf{Pearson correlation coefficient}\\
        \hline
        {} & {SSE} & {SZSE} \\
        \hline
        {random from uniform} & {0.0011922} & {0.0010882} \\
        {random from normal} & {0.00055439} & {0.00050227} \\
        {temperature} & {0.10125} & {0.064122}\\
        {standard sentimental factor} & {0.36284} & {0.37204}\\
        {adjusted sentimental factor} & {\textbf{0.58815}} & {\textbf{0.58042}}\\
        \hline
    \end{tabular}
    \caption{Pearson correlation coefficient from 2015/2/11 to 2015/9/11}\label{Table 5}
\end{table}

From then on, we see that our model can provide a satisfied guide meaning, especially in the special period, namely the period which is influenced greatly by public sentiment.

What's more, we label 2781 of these common words, and just label them as 1(positive) or -1(negative) or 0(indeterminate). After eliminating the words which have computed Senti-score lower than 0.1 and the words with indeterminate sentiment, we find that our model has an accuracy of 73.0088\% on 452 words which has determinate sentiment.

\section{\underline{Conclusion}}

In this paper we develop an algorithm to compute a sentimental factor of Chinese markets and demonstrate that this factor has significant correlation with Chinese market. This factor provides us with a new way to make investment decisions.

The method to compute a sentimental factor is the main contribution of this paper. It will help us even more if we are able to compute sentimental factors for every financial product. Also, a combination of this sentimental factor and traditional financial factors may help us to make even better investment decisions. Looking forward to seeing related research.

\section*{Acknowledgment}

We would like to say thanks to MingWen Liu from ShiningMidas Private Fund for his generous help throughout the research. We are also grateful to Xingyu Fu from Sun Yat-sen University for his guidance and help. With their help, this research has been completed successfully.

\section{Appendix}
\subsection{Figure}

\newpage

\begin{figure}[ht]
  \centering
  \includegraphics[scale=0.4]{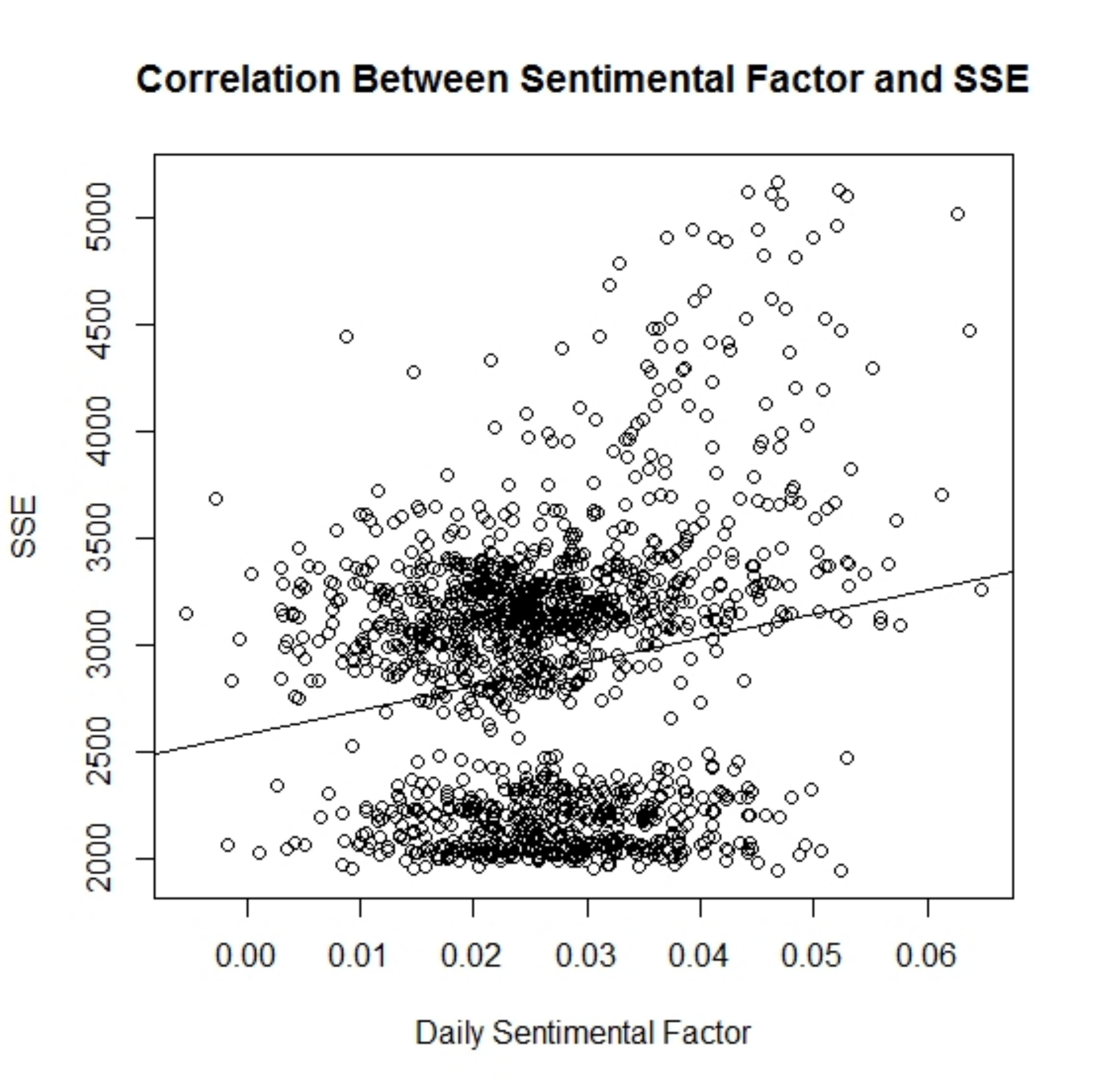}
  \caption{Correlation Between Sentimental Factor an SSE}
\end{figure}

\begin{figure}[ht]
  \centering
  \includegraphics[scale=0.4]{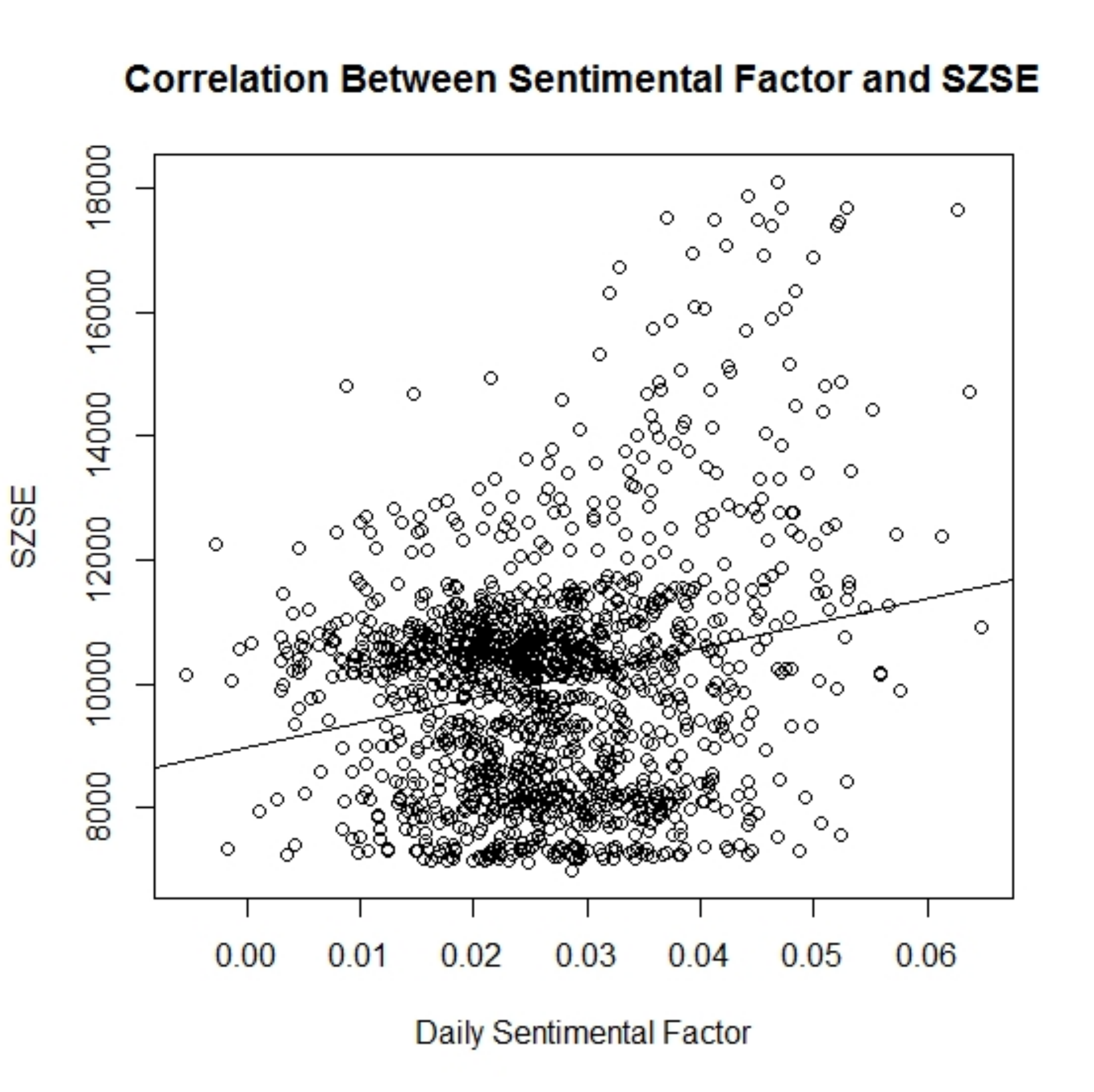}
  \caption{Correlation Between Sentimental Factor an SZSE}
\end{figure}

\begin{figure*}[ht]
  \centering
  \includegraphics[width=\textwidth]{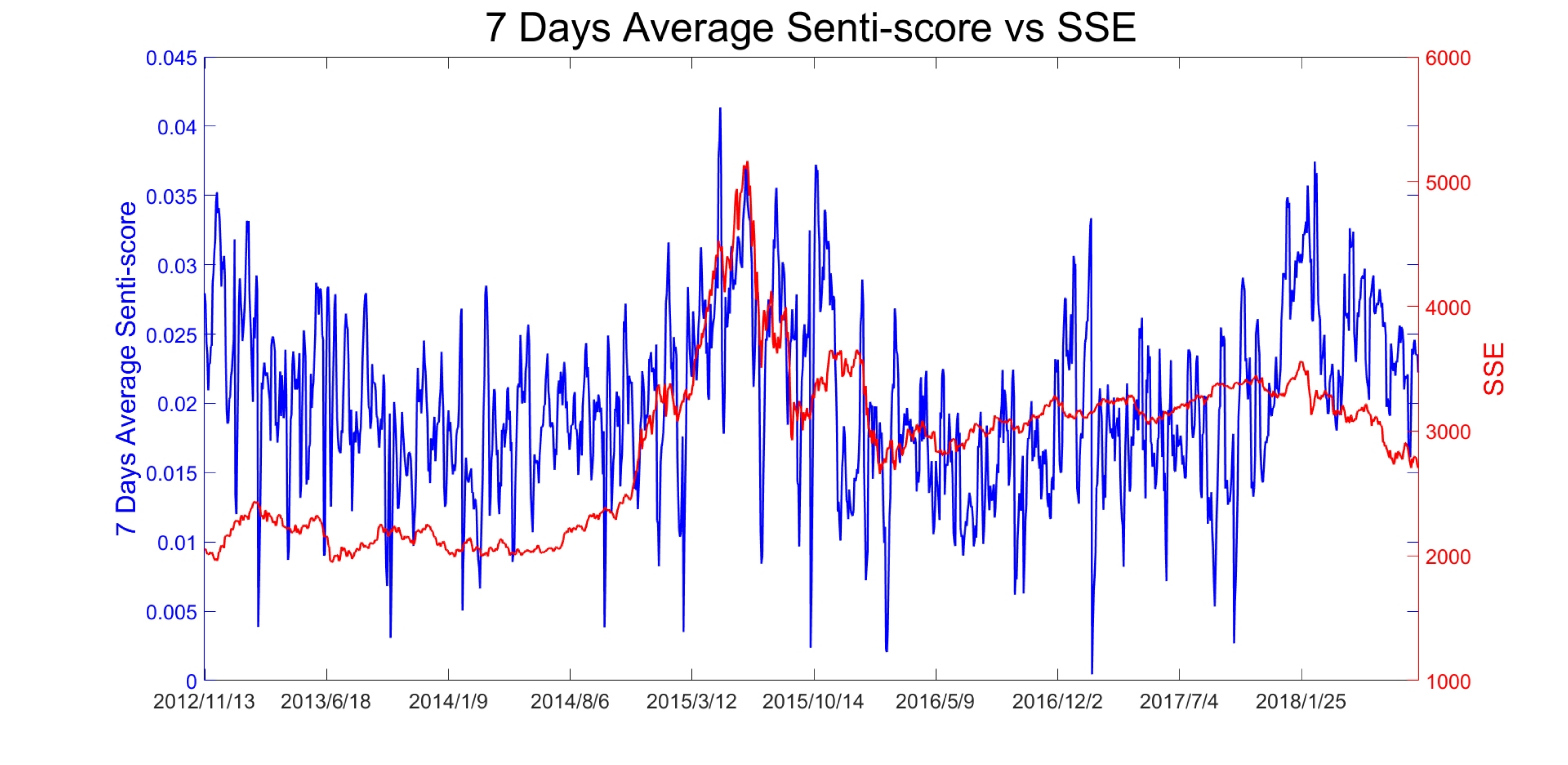}
  \caption{Time Series of 7 Days Average Senti-score and SSE}
\end{figure*}

\begin{figure*}[ht]
  \centering
  \includegraphics[width=\textwidth]{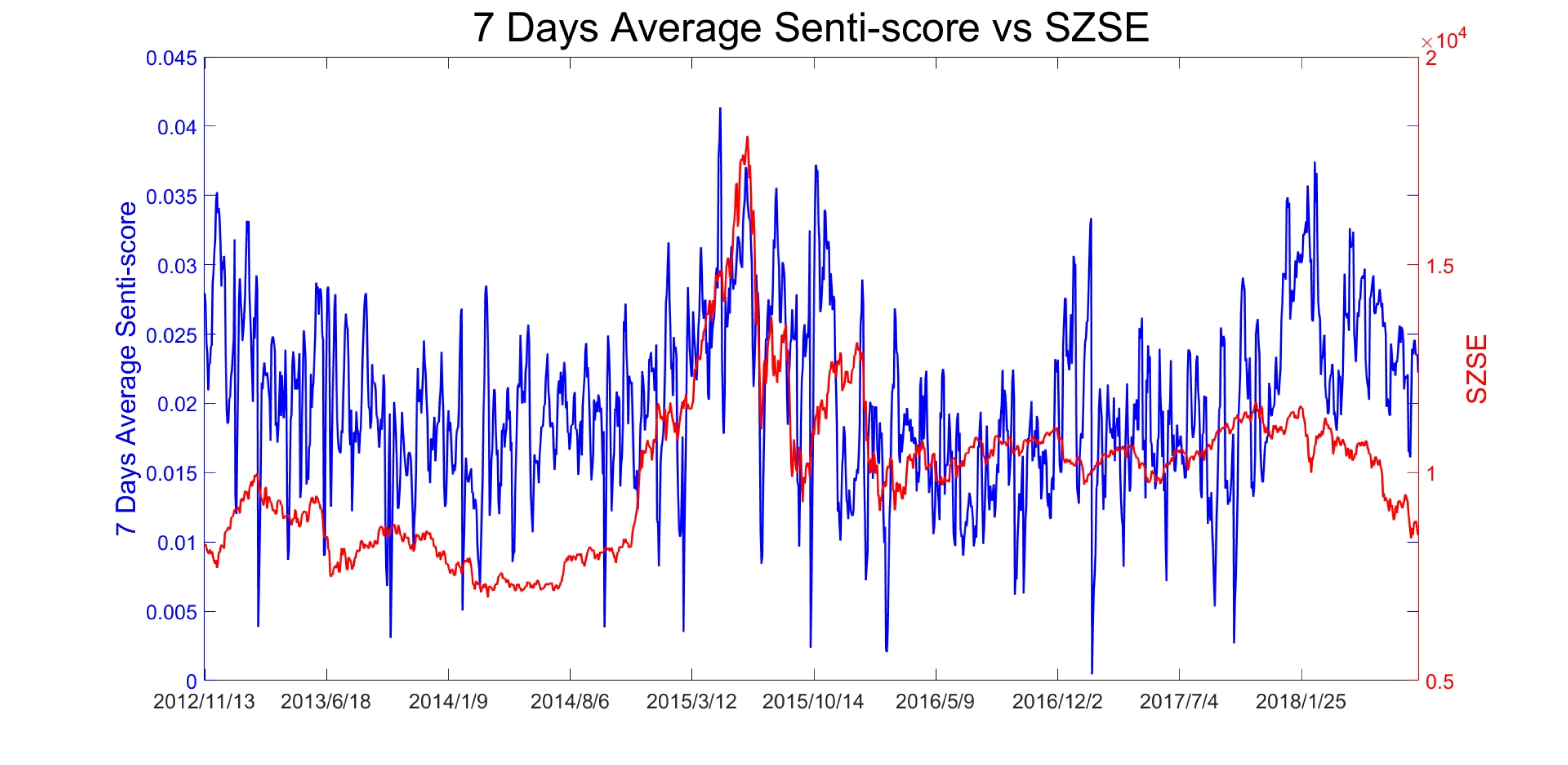}
  \caption{Time Series of 7 Days Average Senti-score and SZSE}
\end{figure*}

\begin{thebibliography}{1}
\bibitem{IEEEhowto:kopka}
Alec Go, Richa Bhayani, Lei Huang, 2009 [R]. Twitter Sentiment Classification Using Distant Supervision.\label{ref 1}
\bibitem{IEEEhowto:kopka}
Mikolov T, Chen K, Corrado G, et al. Efficient Estimation of Word Representations in Vector Space[J]. Computer Science, 2013. \label{ref 2}
\bibitem{IEEEhowto:kopka}
George A. Miller (1995). WordNet: A Lexical Database for English. Communications of the ACM Vol. 38, No. 11: 39-41.\label{ref 3}
\bibitem{IEEEhowto:kopka}
Bird S, Klein E, Loper E. Natural language processing with Python: analyzing text with the natural language toolkit[M]. " O'Reilly Media, Inc.", 2009. \label{ref 4}
\bibitem{IEEEhowto:kopka}
Building the Chinese Wordnet (COW): Starting from Core Synsets. In Proceedings of the 11th Workshop on Asian Language Resources: ALR-2013 a Workshop of The 6th International Joint Conference on Natural Language Processing (IJCNLP-6). Nagoya. pp.10-18. \label{ref 5}
\bibitem{IEEEhowto:kopka}
Mikolov, T., Sutskever, I., Chen, K., Corrado, G. S., \& Dean, J. (2013). Distributed representations of words and phrases and their compositionality. In Advances in neural information processing systems (pp. 3111-3119). \label{ref 6}
\bibitem{IEEEhowto:kopka}
Rao T, Srivastava S. Analyzing Stock Market Movements Using Twitter Sentiment Analysis[C]// International Conference on Advances in Social Networks Analysis and Mining. IEEE Computer Society, 2012:119-123. \label{ref 7}
\bibitem{IEEEhowto:kopka}
Jr E M S. Stock prices and the Wall Street weather[J]. American Economic Review, 1993, 83(3):p��gs. 1337-1345. \label{ref 8}
\end{thebibliography}
\end{document}